\documentclass[12pt]{article}

\usepackage{aaspp4,tighten}

\newcommand{\msolyr}{\ifmmode{{\rm M}_\odot~{\rm 
yr}^{-1}}\else{{M$_\odot$~yr}$^{-1}$}\fi}
\newcommand{\etal}{et~al.}
\newcommand{\msun}{\ifmmode{{\rm M}_{\odot}}\else{M$_{\odot}$}\fi} 
\newcommand{\lsun}{\ifmmode{{\rm L}_{\odot}}\else{L$_{\odot}$}\fi} 

\slugcomment{}

\begin{document}

\title{On the Origin of Planetary Nebula K648 in Globular Cluster M15} 

\author{J.F. Buell, R. B. C. Henry, E. Baron}
\affil{Department of Physics and Astronomy,
University of Oklahoma, Norman, OK 73019 USA}
\begin{center}
and
\end{center}
\author{K. B. Kwitter}
\affil{Department of Astronomy, Williams College, Williamstown, MA
01267 USA}
\begin{abstract}
We examine two scenarios for formation of the planetary nebula
K648: a prompt scenario where the planetary nebula is ejected and
formed immediately after a helium shell flash and a delayed
scenario where a third dredge up occurs and the envelope is ejected
during the following interpulse phase. We present models of both
scenarios and find that each can produce K648-like systems. We
suggest that the prompt scenario is more favorable but cannot rule
out the delayed scenario. 
\end{abstract}

\keywords{globular clusters: individual (M15) --- planetary
nebulae: general --- planetary nebulae: individual (K648) --- stars:
AGB and post-AGB}

\section{Introduction}
The globular cluster M15 contains the well studied planetary nebula
(PN) K648. This is one of the few Galactic PNe with a reasonably
well-determined distance. Therefore, fundamental properties such as
the stellar luminosity can be determined with some confidence. Because
of its globular cluster membership, many of the progenitor properties,
such as the zero age main sequence (ZAMS) mass, can be inferred
reliably.

Due to the importance of K648 as a halo PN, it has been the focus
of several abundance studies, and all of these show the abundances of
most metals to be depleted relative to the sun, consistent with a
progenitor of low metallicity.  Carbon is an exception; studies which
determine the ratio (by number) of C/O in K648 infer values that range
from $4-11$ (Adams \etal\ 1984; Henry, Kwitter, and Howard 1996;
Howard, Henry, and McCartney 1997), which is far above C/O in the Sun
of 0.43 (\markcite{ag89}Anders and Grevesse 1989, hereafter
AG89). This is in fact much higher than the average C/O ratio of
$\approx 0.8$ for solar neighborhood PNe (\markcite{rs94}Rola and
Stasi{\'n}ska 1994).

Low and intermediate mass stars that have left the main sequence,
ascended the giant branch, and passed through the horizontal branch, 
then enter a thermally unstable phase where energy is generated by
shell He and H-burning called the thermally pulsing asymtoptic giant
branch (TP-AGB) stage, which is a very important yet not well understood
phase [Detailed reviews of this stage can be found in \markcite{i95} 
Iben (1995), \markcite{l93} Lattanzio (1993), and \markcite{ir83} Iben
and Renzini (1983)]. During the TP-AGB stage the star alternates
between a long stage where the luminosity is generated mostly by
quiescent hydrogen shell burning, with a helium burning layer
producing a minority of the energy, and a thermal runaway stage in the
unstable helium burning layer (\markcite{sh65} Schwarzschild and
H\"{a}rm 1965, 1967; and \markcite{w66} Weigert 1966).  The second
stage results in expansion of the outer layers and an extinguishing of
the H burning shell. This short stage, characterized by rapid changes,
with helium burning dominating the energy generation, is known as a
thermal pulse or a He shell flash.

TP-AGB stars exhibit large mass-loss rates ranging from
10$^{-7}-$10$^{-4}$ \msolyr\ . Indeed such high mass-loss rates
are predicted to result in the ejection of the envelope, at which
point the star leaves the AGB and becomes a planetary nebula central 
star (CSPN). The first models of CSPN tracks were made by
\markcite{p91} Paczynski (1971) who showed that the CSPNs evolve
horizontally on the HR diagram when nuclear burning is still taking
place and then as they cool the luminosity and temperature
decrease. \markcite{hs75} H\"arm and Schwarzschild (1975) showed that 
a CSPN could leave the AGB as either a helium burning or hydrogen
burning star. The observational consequences of hydrogen and helium
burning have been studied in the more refined models
including mass loss showed that the subsequent evolution of the
central star depends on whether the star leaves the AGB as a
helium or hydrogen burner [Sch\"{o}nberner (1981, 1983) and
\markcite{i84} Iben (1984)]. 

Low mass stars (M$\lesssim$ 3 M$_{\sun}$) can experience two mixing
episodes or ``dredge-ups''. During dredge-up, material that has been
processed by nuclear burning is mixed into the surface layers. At the
entrance to the giant branch, the convective region can extend into
the core, leading to mixing of CNO products into the outer
layers. Similarly as shown by \markcite{i75} Iben (1975), after a
thermal pulse on the AGB, the convective region can extend into the
core, mixing He-burning products into the outer layers. These two
mixing events are known as first and third dredge up, respectively
(second dredge up will not concern us here). Therefore, a third
dredge-up is a natural explanation of the high carbon abundance found
in K648.  On the other hand, no carbon stars have been observed either
in M15 or in any other globular cluster, although such stars should be
the immediate progenitors of objects such as K648 if a third dredge-up
occurs. Thus, the lack of carbon stars in M15 weakens the argument for
a third dredge-up event.

One possible explanation for the absence of carbon stars is a delayed
scenario in which the third dredge-up of carbon rich material changes
the structure of the envelope during the following interpulse phase,
ultimately increasing the mass-loss rate significantly and driving off
the stellar envelope (Iben 1995). Thus envelope ejection is delayed
until the interpulse phase following this dredge-up of carbon rich
material. 

Another explanation supposes that the
envelope is removed during the quiescent He-burning stage that follows
a thermal pulse (\markcite{ren89}Renzini 1989 and
\markcite{rfp88}Renzini and Fuci-Pecci 1988).  The carbon then
originates in a fast wind from the central star (CSPN).  In addition,
the wind produces shock-heating in the nebula, which, if not properly
accounted for during an abundance analysis, may lead to the inference
of a spuriously high C/O ratio. In this case the envelope would be
ejected immediately after a thermal pulse while helium shell burning
still dominates the luminosity.  We refer to this mechanism as the
prompt scenario. 

In this paper we calculate detailed envelope models of thermally pulsing
asymptotic giant branch star envelopes to test the predictions of the
delayed mechanism, perform other calculations relevant to the prompt
mechanism, and compare output of each with observations of K648.
Section~2 describes the envelope code, section~3 presents the
observational data and the results for the delayed and prompt models,
and a brief discussion of our findings is given in section~4.

\section{Models}
The computer code used to calculate the delayed models is a
significantly updated and modified version of a program kindly
provided to us by \markcite{ren92} A.~Renzini for modeling
the envelope of  TP-AGB
stars during the interpulse phase.  Many of the basic details of the
method are enumerated in \markcite{it78} Iben and Truran (1978) and
\markcite{rv81} Renzini and Voli (1981) and references therein; in
this section we concentrate on those features which are different. In
a future paper (\markcite{buell97}Buell \etal\ 1997) we will provide
a more detailed description of the code. 

The mass of the hydrogen exhausted core ($M_H$) at the first thermal
pulse is given by the expression found in \markcite{latt86}Lattanzio
(1986).  During each interpulse phase the code follows the mass of the
hydrogen exhausted core and envelope, the evolution of envelope
abundances of $^4He$, $^{12}C$, $^{13}C$, $^{14}N$, and $^{16}O$, and
determines $T_{\rm eff}$ by integrating the equations of stellar
structure from the surface to the core.
Envelope abundances at the first pulse are determined by combining
published main sequence levels with changes due to the first
dredge-up. The former are established by scaling the AG89 solar
abundances of all metals except the alpha elements, i.e. oxygen, neon,
and magnesium, to the appropriate metallicity, and then setting
$[N_{\alpha}/Fe]=0.4$, where $N_{\alpha}$ is the number abundance of
O, Ne, and Mg. This last value is chosen from an examination of the
trends in the data of \markcite{edv93} Edvardsson \etal\ (1993) for
[Fe/H]$<$-1.0 and by assuming that neon and oxygen vary in lockstep in
PNe as shown by \markcite{hen89}Henry (1989). The abundance changes
due to the first dredge-up are calculated from the formulae of
\markcite{gwdj93}Groenewegen and deJong (1993). 

The mass-loss both before and during the TP-AGB phase is very
important, although the parameters are poorly understood. The
pre-TP-AGB mass-loss is a free parameter, while during the TP-AGB
phase, mass-loss is determined by using the expression of
\markcite{vw93}Vassiliadis and Wood (1993), which can be written as
\begin{eqnarray}
\log{\dot{M}=-11.43+1.0467\times
10^{-4}\left(\frac{R}{R_{\sun}}\right)^{1.94}
\left(\frac{M}{M_{\sun}}\right)^{-0.9}}& {\msolyr}.
\end{eqnarray}
The above rate is used until $\log{\dot{M}} = -4.5$, and then it is
held fixed. Equation 1 is a $\dot{M}\rm -Period$ relation based on
mass-loss from population I stars.  However, recent calculations by
\markcite{wbs95} Wilson, Bowen, and Struck (1995) suggest that the
mass-loss rates of low metallicity AGB stars are also strongly
dependent on radius. There is considerable uncertainty in this
equation.  For example, predicted mass-loss rates from other equations
with a similar form (e.g. Bazan 1991) differ from predictions of
eq.~(1) by up to a factor of five.

The luminosity of TP-AGB stars after the first few pulses can be
described by a linear relation between core-mass and luminosity as
first discovered by \markcite{p90} Paczynski (1970). Models of TP-AGB
stars have shown that for  $M\lesssim3.0\msun$ this
relation depends on metallicity (Lattanzio 1986, Hollowell and Iben
1988, Boothroyd and Sackmann 1988b). At the first pulse the luminosity
of TP-AGB stars is less than the asymptotic core-mass-luminosity
relation. The luminosity at the first pulse in our models is found by
linearly extrapolating in metallicity from the expressions found in
\markcite{bs88b} Boothroyd and Sackmann (1988b).  After the first
pulse, the luminosity of the AGB star rises steeply until it reaches a
value predicted by the core-mass luminosity relation (CML) of
\markcite{bs88b}Boothroyd and Sackmann (1988b).  This relation
predicts luminosity primarily as a function of core mass, although it
has a weak dependence on helium and metal mass fractions.
 

Carbon rich material can be dredged from the core into the envelope
following a thermal pulse. We assume that when the mass of the
hydrogen-exhausted core exceeds a minimum mass ($\rm M^{DU}_{min}$)
that a dredge-up occurs. The amount of material dredged up, $\Delta
\rm M_{dredge}$, is determined by the free parameter $\lambda$, where
\begin{equation}
\lambda=\frac{\Delta \rm M_{dredge}}{\Delta M_c}.
\end{equation}
In eq.~(2) $\Delta M_c$ is the amount of core advance during the
preceding interpulse phase. We determine the composition of the
dredged up material from the formulas in \markcite{rv81} Renzini and
Voli (1981), with $^4He\approx 0.75$, $^{12}C\approx 0.23$, and
$^{16}O\approx 0.01$ as the approximate mass fractions.

Finally, the code uses the opacities of \markcite{opal} Rogers and Iglesias
(1992) supplemented by the low temperature opacities of
\markcite{af94} Alexander and Ferguson (1994). 

\section{Results and Discussion}
\subsection{Observational Parameters}
Numerous observed and inferred parameters for K648 are listed in Tables
1a and 1b, where the symbols in column~(1) are explained in the table
notes.  We comment here on the method of determination for several of
them. 

Radio images of K648 have been made by Gathier \etal\
(1983) \markcite{gpg83} and optical images were made by
\markcite{adams84} Adams \etal\ (1984), and recently by
\markcite{bianchi95} Bianchi \etal\ (1995) using the HST. The HST data
called into question the small size for the nebula inferred in the
radio studies of Gathier \etal\ (1983) \markcite{gpg83} and the
optical studies of Adams \etal\ (1984)\markcite{adams84}, since HST
was able to resolve the structure of the nebula. This leads to, e.g.,
a larger planetary mass and smaller electron density. In Tables 1a and
1b, we quote all results.

M$_{PN}$ was computed using equation V-7 in \markcite{pott84} Pottasch
(1984), while the dynamical age was estimated by dividing the nebular
radius by the expansion velocity ($v_{exp}$). Since no $v_{exp}$ is
available for K648 we use a range which represents typical values for
PNe. The central star mass for K648 was estimated by linearly
interpolating/extrapolating using both hydrogen burning and helium 
burning post-AGB tracks of \markcite{vw94} Vassiliadis and Wood (1994)
in the log~L-log~T plane.  The metallicity of M15 suggests using a low
Z track, although the carbon abundance of K648, if correct, would
increase the metallicity of the star, suggesting that a higher Z track
is more appropriate. Since the metallicity dependence is unclear, we
estimated the range of possible central star masses by performing the
interpolation for each metallicity considered by Vassiliadis and Wood.
Thus, the mass range of the central star is 0.55--0.58$M_{\sun}$ for
the H burning tracks and 0.56--0.61$M_{\sun}$ for the He-burning
tracks.  We adopt a final core mass of $0.58\pm0.03M_{\sun}$.

The theoretical age of the central star was estimated from the figures
of Vassiliadis and Wood and 
linearly interpolating in $\log{\rm L}$ between
tracks which closely match the core mass of K648, e.g., the hydrogen
burning M$_{c}$=0.56, Z=0.016 track and the helium burning
M$_{c}$=0.56 M$_{\sun}$, Z=0.004 give evolutionary ages of $\sim$
12000~yr and $\sim$1800~yr, respectively. Other tracks with M$_{\rm
c}\lesssim 0.6\msun$ and different metallicities give similar
results. When compared to the dynamical age a He burning track is
favored. 

The adopted abundances of K648 for He/H, C/O, and N/O ratios represent
a range of recent literature values.  Howard \etal\ (1997) find that
in six of the nine halo PNe they studied, the C/O ratio exceeds the
solar value.  Many of these nebulae have stellar temperatures much
higher than that of K648, implying that they are older and more
evolved. Since the high C/O ratios persist into the later stages of PN
evolution, this suggests that the inferred C/O is not influenced by
the presence of shock heating in the nebula.

The mass-loss rate at the tip of
the AGB was determined by dividing the nebular mass by the dynamical
age. This is in reality a lower limit since it assumes that the nebula
has a filling factor of 1, which is unrealistic. By this procedure we
calculate that the lower limit to the mass-loss is 9$\times$10$^{-6}$
\msolyr . The upper limit is assumed to be $10^{-4}$ \msolyr .

The composition of the central star is uncertain, as two recent papers
do not agree. McCarthey \etal (1996) find that the central star has a
normal helium abundance, whereas Heber \etal (1993) find that the central
star is helium and carbon rich.

\subsection{Delayed Scenario}

We have calculated several low mass, low metallicity models, but here
we focus on the two models listed in Table 2, where we present the
model input parameters: the ZAMS mass (M), the core mass at PN
ejection (M$_c$), the mass of the PN ($M_{PN}$), the ZAMS [Fe/H]
ratio, the adopted ratio of the mixing length to pressure scale height
($\alpha$), the mass of the model star at the first thermal pulse 
(M$_{FTP}$), the adopted dredge-up parameter ($\lambda$), and the
minimum core mass for dredge-up (M$_{c,min}^{DU}$).  The panels of
Figure 1 show the evolution of the interpulse luminosity, the stellar
radius, the mass-loss rate, and the core mass as a function of total
mass. All quantities are expressed in solar units. Figure 2 shows the
evolution of the chemical composition of the envelope as a function of
total mass. Table~1b compares the observed quantities to our
predicted ones.

We note in Figure~1 that the interpulse radius of each model star
increases dramatically after the final pulse, as compared to the
preceding interpulse phase. The increase in radius leads to a large
increase in the mass-loss rate in each model during the final
interpulse phase; the mass-loss rate increases by almost a factor of
100 in model 2 and by a factor of 5 in model 1. This is a consequence
of the steep dependence of our mass-loss law on the stellar
radius. The significant increase in the mass-loss rate causes the star
to lose its envelope in a few thousand years. The mass-loss rate for
model~1 is clearly too small relative to the observationally derived
value.  However, we have found that by reducing the mixing length
($\alpha$) by a factor of two, as we have done in model 2, we can make
a model that essentially reproduces the observed AGB tip mass loss
rate.

The significant event that occurs during the final pulse is a
dredge-up of helium and carbon rich material. The mass of material
dredged up is a few times $10^{-5}M_{\sun}$. However, given the mass
of the envelope and the low initial abundances, the amount of carbon
dredged into the envelope is significant enough to increase the carbon
mass fraction by a large factor in each case.  Consequently, the
envelope opacity rises, causing a dramatic increase in the stellar
radius.

The envelope of each model at the last thermal pulse is only a few
times $10^{-2} M_{\sun}$ and, after the final carbon dredging pulse,
is ejected on a timescale of a few hundred years. Each model star is a
carbon star for only a few hundred years, due to the rapid mass-loss
after a dredge-up of carbon. 
This short lifetime, coupled with the relatively low incidence of PNe
in globular clusters [two confirmed and three possible candidates
(Jacoby \etal\ 1995)], perhaps explains why carbon stars have not been
observed in globular clusters.


An important check on our models is to compare the predicted AGB tip
luminosity with its observed value. The predicted luminosity of our
models at the top of the AGB agrees fairly well with the tip of M15's
red giant branch (Adams \etal\ 1984\markcite{adams84}). Our models
suggest that the observed AGB tip will actually correspond to the
second-to-last pulse, since after the dredge-up event the star is
predicted to remain as an AGB star for only $\sim$1000~yr. The
luminosity of K648 in Adams \etal\ (1984) appears to be 0.1dex higher
than the tip of the giant branch, this may be due to the metallicity
enhancement due to the dredge up. As noted earlier, the core-mass
luminosity relationship depends on metallicity, with the luminosity at
a set core-mass increasing with increasing metallicity. If we lower
the luminosity still further to $\sim$2000\lsun\ to match the tip of
the giant branch, we believe that the addition of carbon 
to the envelope will still cause envelope ejection.

There is some question about whether or not dredge-up can occur at the
low values of M$_{c,min}^{DU}$ indicated by our models (see
Table~2). While \markcite{latt89}Lattanzio (1989) found that dredge-up
can occur at a core mass above $0.605M_{\sun}$, the same study also
found a dependence of the minimum dredge-up mass on metallicity, with
lower metallicities giving lower mass dredge-ups. \markcite{bs88c}
Boothroyd and Sackmann (1988c) found that if they increased the mixing
length parameter $\alpha$ from 1 to 3, they were able to cause a
dredge-up in a model with Z=0.001, $\rm M_c=0.566M_{\sun}$, and $\rm
M=0.81M_{\sun}$, although it is unclear if a mixing length this large
is justified. Additionally, to match the low luminosity end of the
carbon star luminosity function of the LMC, Groenewegen and deJong
(1993) had to set $\rm M^{DU}_{min}=0.58M_{\sun}$. From these studies
it appears that our values for M$_{c,min}^{DU}$ are not unreasonable.

Finally, we point out that each of the delayed models gives a very
natural explanation of the high carbon abundance of K648 and the lack
of carbon stars. Each model also predicts the observed mass of the
ionized gas to be a few times $10^{-2}M_{\sun}$. The C/O ratios of
each model range from 4 to 25, with model 2 giving the best fit, which
agrees reasonably well with the observed values of $4-11$. The He/H
ratios of the model stars also agree with the observed value of
0.09. The high N/O ratio inferred in the models may be an artifact of
our choice of initial O abundance and hence could be reduced with a
higher O abundance, which would also slightly reduce the C/O ratio.
Thus, our delayed models are consistent with several important
observed properties of the K648 system.

\subsection{Prompt Scenario}

An alternative scenario results if we apply our mass-loss formulation
to the secondary luminosity peak (SLP) which follows the helium shell
flash of the $\rm 1M_{\sun}$, Z=0.001 model of \markcite{bs88a}
Boothroyd and Sackmann (1988a, BS88a). The metallicity of the BS88a
model is a factor of $\sim$5 higher than M15, however, no models of
the appropriate metallicity exist and we attempted to use the closest
one in terms of Z, M$_{\rm c}$, and M. The SLP corresponds to the
region between point C and the vertical dashed line on figure 2 of
Boothroyd and Sackmann (1988a), i.e. the same place that Renzini (1989) and
Renzini and Fuci-Pecci (1988) predict this event to occur when the
star expands. The SLP occurs when the excess luminosity produced in
the helium shell flash reaches the surface. This peak can be seen in
most models of low mass AGB stars [Iben(1982), BS88a, VW93]. 

It should be noted that this is the
point where dredge-up can occur, although it does not necessarily
do so. This scenario does not require a dredge-up of carbon rich
material for envelope ejection. We define the prompt scenario as
ejection at the SLP without the dredge-up of carbon rich material.

The adopted parameters of this model are shown in Table
3. The luminosity and radius are eyeballed lower limits from the
SLP of BS88a, while the mass and core mass are parameters stated in
their text. The mass-loss rate calculated from our prescription
[i.e. eq.~(1)] is $\sim 10^{-5}\msolyr$ (essentially the Eddington
limit), which will remove the $\rm 0.03M_{\sun}$ envelope in a few
thousand years. This model is similar to K648 in terms of core mass
and envelope mass.  Values for luminosity, radius, mass-loss rate, and
core mass for the prompt scenario are indicated with filled diamonds
in Fig.~1 and observed quantities are also compared to those predicted
for this scenario in Table~1b. 

A carbon-rich nebula could be formed by the prompt mechanism if a
sufficient amount of helium and carbon-rich material is ejected during
the post-AGB phase and mixed with the ejected hydrogen-rich
envelope. The fast wind overtaking the slower wind will produce a
shock which would likely be Rayleigh-Taylor unstable, causing the
nebula to mix. Only $5-15\times$10$^{-5}$ M$_{\sun}$ of material with
mass fractions of $^{4}He$=0.75 and $^{12}$C=0.23 needs to be mixed
into the envelope to match the C/O ratio of K648. Carbon-rich material
can be ejected into the nebula during the post AGB phase. As a star
moves horizontally across the HR diagram from the AGB stage to CSPN
position, the mass-loss rate will decrease as the wind speed
increases, so the material ejected during this transition can be mixed
with the slower hydrogen rich envelope. And since the currently
observed mass-loss rate of the K648 central star is
$10^{-9}-10^{-10}$~\msolyr\ (Adams \etal\ 1984; Bianchi \etal\ 1995),
the nebula is no longer being polluted. Examination of the models of
Vassiliadis and Wood (1994) suggests that as the star moves from the
AGB phase to the CSPN phase, the mass loss rate drops from
10$^{-5}$~\msolyr\ to $\sim10^{-10}$~\msolyr, indicating that during
this transition the mass loss rate was higher in the past, and possibly
high enough to account for the carbon enrichments in K648.

In the prompt scenario, the envelope is ejected when the star is
burning helium and as a result the resulting CSPN will
be follow a helium burning track (Sch\"{o}nberner 1981, 1983;
Iben 1984).



Thus, in the prompt scenario, the evolved star ejects sufficient
carbon into a slower moving hydrogen-rich shell to produce the PN we
observe today.  Mixing is assumed to occur due to shock induced
instabilities. Since the prompt scenario postulates the removal of the
entire H-rich envelope during the He burning stage, we expect K648 to
follow a He burning track because the H-burning shell has been
extinguished during the thermal pulse. Ultimately, a white
dwarf of type DB will be produced. 

\section{Discussion}
One additional scenario is again a delayed one, but one in which the
CSPN is a helium burner. We have not as yet performed calculations
relevant to it. In this case, if a dredge-up occurs, it does so at the
SLP. The stellar envelope will be enriched in carbon and the added
opacity may allow an even greater expansion during the SLP, making it
more likely that the envelope will be ejected during this phase. The
resulting PN would be carbon rich and have a helium burning CSPN. We
feel that this is also a promising model, although, proper calculations
of this scenario need to be done.

Both the prompt and delayed scenarios can be made to match many of the
observed features of K648. With each mechanism the radius increases
dramatically: in the prompt because of the increase in luminosity of
the star after a thermal pulse and in the delayed because of an
increase in the opacity due to an infusion of carbon rich material.
In addition, both mechanisms produce $^{12}$C in sufficient amounts to
explain the observed C/O ratio.
 
The most serious difficulty with the prompt scenario is that it
can only explain the enhancement of the carbon and helium abundances
by essentially adhoc means, in this case assuming the central star
wind pollutes the rest of the nebula or by shocks and carbon-rich
pockets due to this wind. This may not be an unreasonable assumption,
since the mass of K648 is low compared to a ``typical'' PN
($\sim$0.1M$_{\sun}$). To test the prompt scenario would require a
detailed model following the star from the horizontal branch to the
central star phase with attention to the details of mass-loss to see
if the central star wind can truly enhance the carbon and helium
abundances of the PN and multidimensional hydrodynamics to test the mixing
hypothesis. 

The difficulty with the delayed scenario is it predicts that the
CSPN should be a H-burner. The dynamical age of K648 favors a
He-burning CSPN which is more likely to occur in the prompt 
scenario as the envelope is ejected during a phase when helium burning
is dominant. Since we assume that for a given metallicity only one of
these scenarios will be operative, a strong observational test to
determine the correct scenario would be to search for white dwarfs in
M15. If they are found to be type DB, this would favor the prompt
scenario, and if they are type DA, the delayed scenario is more
likely. 

A point favoring the prompt scenario is that it naturally accounts for the
dynamical age. On the other hand, this scenario requires the
assumption of efficient mixing and there is some evidence
(cf. section~3.1) that signatures of the requisite shocks are not
actually observed. However, until detailed models are produced, both
remain as viable evolutionary scenarios for K648 and similar systems.

\acknowledgments
We thank the anonymous referee for pointing out some missing
references in the original manuscript. This work was supported in part
by NASA grant NAG 5-2389 and by NSF grant AST-9417242.

\clearpage

\figcaption[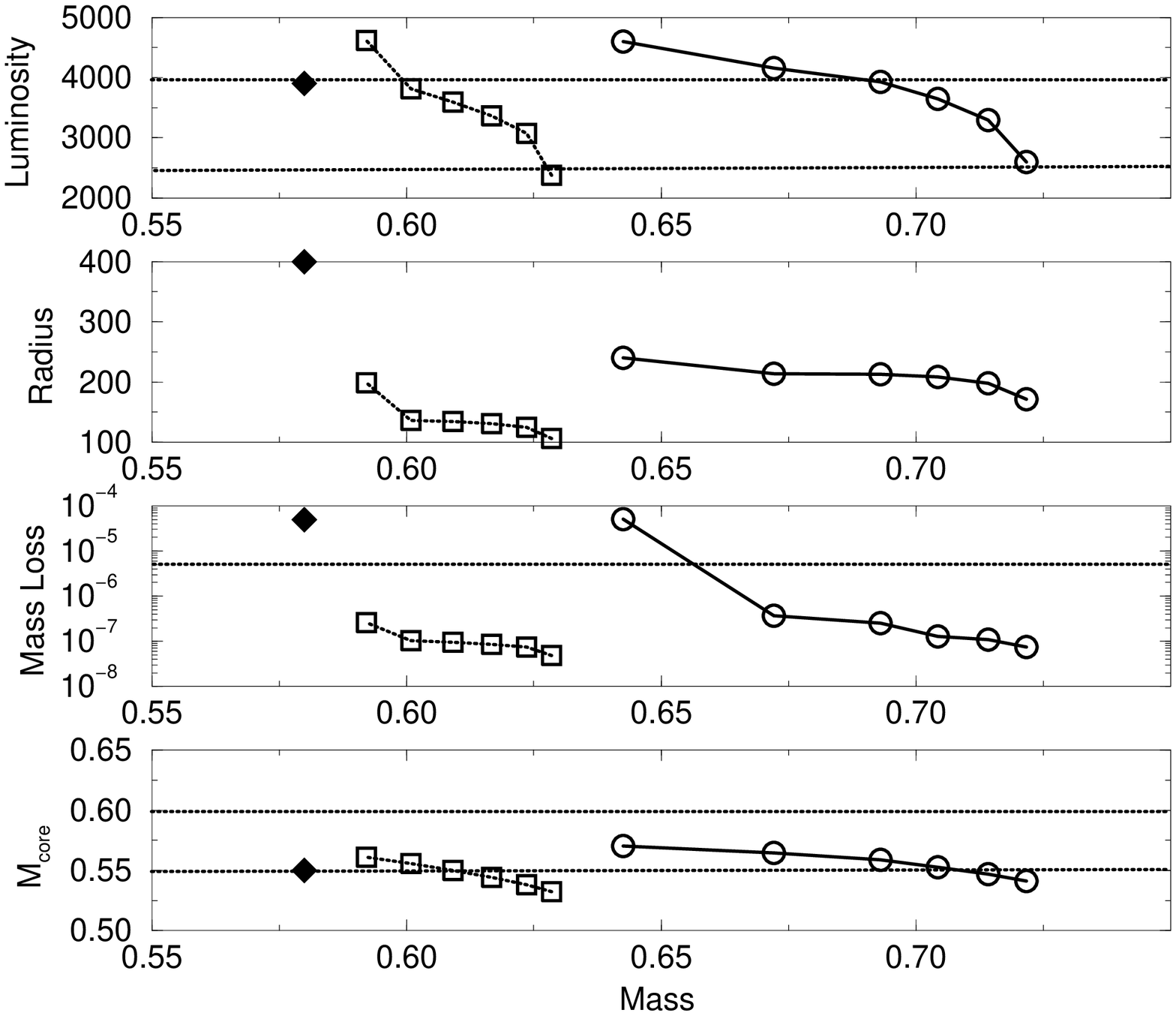]{Shown in the panels of this figure are the
evolution of our thermally pulsing AGB models and the parameters of
our prompt model. The dotted line and open squares track model 1,
the solid line and open circles track model 2, and the solid diamonds
are the parameters of the prompt model. The abscissa of the graphs 
tracks the mass of the models in solar masses. Due to  mass loss the
stars move from right to left on the graphs. The parameters in the
panels are for the interpulse phase. From top to bottom the parameters
are stellar luminosity (in L$_{\sun}$), radius (in $\rm R_{\sun}$),
the mass-loss rate (in \msolyr), and the mass of the core. The
observed upper and lower limits of the AGB tip luminosity are
indicated with dark long dashed lines, the lower limit on the AGB tip
mass-loss rate is indicated with a long dashed line, and the upper and
lower limits of the central star mass are indicated with long dashed
lines.}  

\figcaption[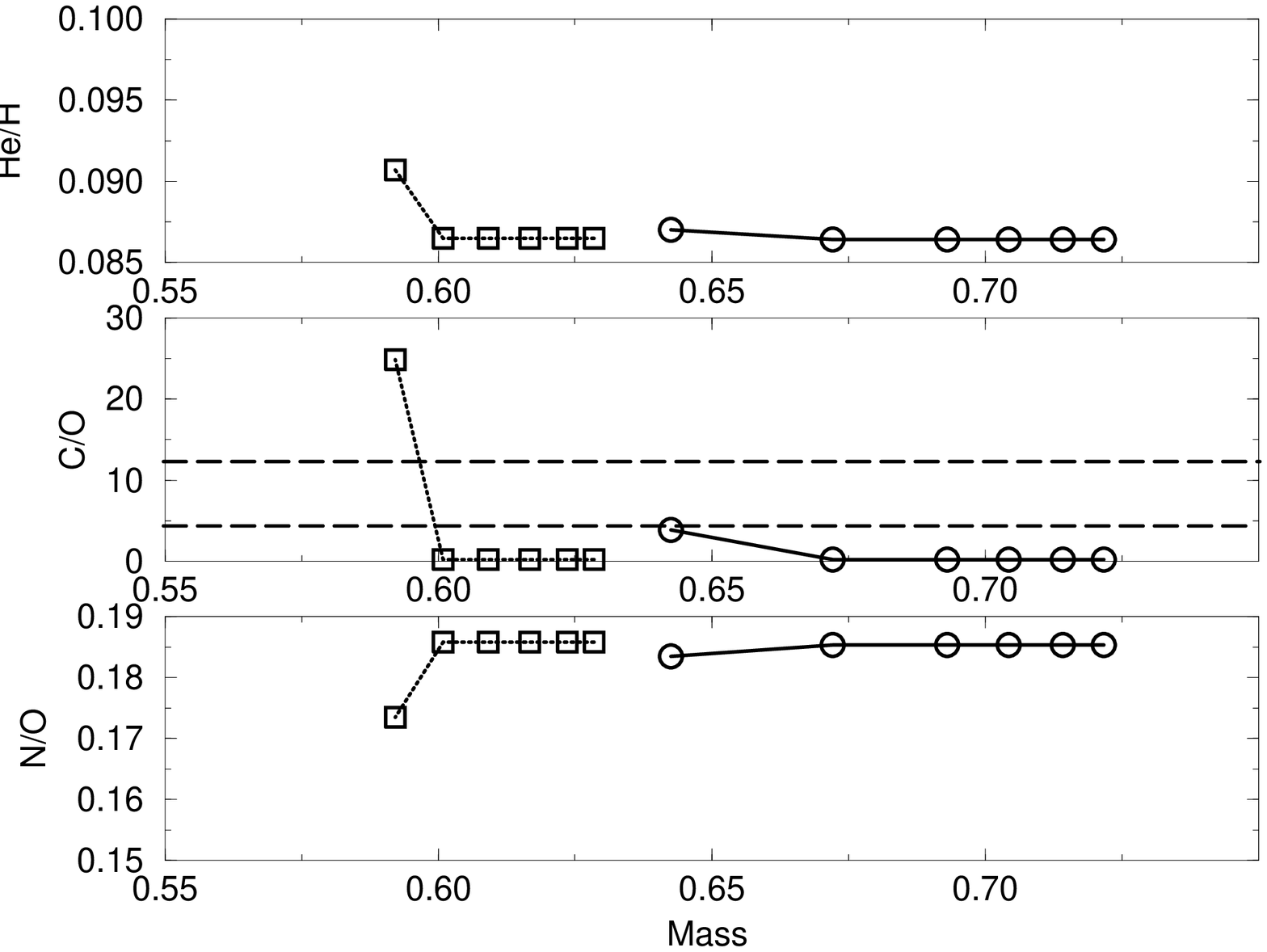]{Shown in the panels of this figure is the
evolution of the surface abundance ratios. The symbols have the same
meaning as the first figure. In the C/O panel the upper
and lower observational limits are shown on the figure with the dark
long dashed line. The range of possible He/H and N/O is encompassed by
the ordinates of these figures.}

\clearpage

\begin{deluxetable}{ccc}
\small
\tablenum{1a}
\tablewidth{33pc}
\tablecaption{Observational Data for K648}
\tablehead{\colhead{Parameter} &\colhead{Value} &\colhead{ref.}} 
\startdata
T$_{\rm eff}$& $36000\pm 4000{\rm\ K}$ & 1,2\nl
d& $10.0\pm 0.8{\rm\ kpc}$ & 3\nl
$\theta$& $1.0-2.5\arcsec$ &1,2,5\nl
$v_{exp}$& $15-25{\rm\ km~s}^{-1}$& \nl
$\rm n_e$& $1700-8000{\rm\ cm}^{-3}$ & 1,2\nl
$\rm T_{e}$& 12000{\rm\ K}& 1,2\nl
$\log{F_{H\beta}}$& $-12.10\pm 0.03$ & 4\nl
\enddata
\tablerefs{ (1) Adams \etal\ 1984; (2) Bianchi \etal\ 1995; (3) Durell
and Harris 1993; (4) Acker \etal\ 1992; (5) Gathier \etal\ 1983}
\tablecomments{This table is a summary of the observed and
inferred parameters for PN K648. The effective
temperature, T$_{\rm eff}$, refers to the central star, while the distance,
d, is the adopted distance to K648. The following nebular parameters
are also listed: the angular size of the nebula, $\theta$; the
expansion velocity, $v_{\rm exp}$; the electron density, $\rm n_e$;
the ionized gas temperature, $\rm T_{e}$; and the
log of the measured H$\beta$ flux in erg~cm$^{-2}$~s$^{-1}$. The large
range in $\theta$ and $n_e$ arise from differences between newer HST
data and ground based data, The HST data give higher a
value of $\theta$ and a lower  value for $n_e$.}
\end{deluxetable}

\clearpage
\begin{deluxetable}{ccccc}
\small
\tablenum{1b}
\tablewidth{33pc}
\tablecaption{Observational Data and Models Compared}
\tablehead{\colhead{Parameter} &\colhead{Observed Value}
&\colhead{ref.}& \colhead{Delayed Scenario}& \colhead{Prompt Scenario}}
\startdata
L& $3200-4700\ L_{\sun}$ & 1,2 & 4600 & 4000\nl
$\rm M_{PN}$ & $0.015-0.090\ M_{\sun}$& 1,2&$0.048\pm 0.012\ M_{\sun}$&$0.064\
M_{\sun}$ \nl
$\rm M_c$ & $0.58 \pm 0.03$ & & $0.57 \pm 0.01$ & $0.58$ \nl
$\tau _{dyn}$& $2000-8000{\rm\ yr}$& &12000{\rm\ yr}& 1800\ yr\nl
He/H& $0.083-0.10$& 1,3,4 & $0.087-0.091$ & $0.9$ \nl
C/O& $4-11$& 1,3,4& $4-25$& $4$ \nl
N/O& $0.05-0.20$&1,3,4& $0.17-0.19$& $0.17$\nl
\enddata
\tablerefs{ (1) Adams \etal\ 1984; (2) Bianchi \etal\ 1995; (3) Henry,
Kwitter, and Howard 1996; (4) Howard, Henry, and McCartney 1997}
\tablecomments{This table compares  the observed and predicted parameters
for PN K648. The observed luminosity, L, refers to the central star,
while the predicted luminosity is the luminosity on the AGB, but since
the tracks are nearly horizontal they should be comparable. The
following nebular parameters are also listed:   the mass of ionized
gas in the nebula, M$_{PN}$; the mass of the central star, $M_c$; the
dynamic timescale, $\tau_{dyn}$; and the abundance ratios He/H, C/O,
and N/O by number. The abundances for the prompt scenario are
calculated assuming $0.00014M_{\sun}$ of helium and carbon rich
material is removed by mass-loss from the CSPN. The observed value for
the dynamical timescale, $\tau_{dyn}$, corresponds to an upper limit
for the age of the nebula.  The theoretical values correspond to
evolutionary time scales required to reach a given central star
temperature. 
The large
range in $L$ and $M_{\rm PN}$ arise from differences between HST
data and ground based radio and optical data.  The HST data give higher values for $L$ and $M_{\rm PN}$.}
\end{deluxetable}

\clearpage

\begin{deluxetable}{ccccccccc}
\tablenum{2}
\tablewidth{33pc}
\tablecaption{Input Parameters and Results for Delayed Models}
\tablehead{\colhead{No.} &\colhead{M} &\colhead{$\rm
M_c$} 
&\colhead{$\rm M_{PN}$} &\colhead{[Fe/H]} 
&\colhead{$\alpha$} &\colhead{$\rm M_{FTP}$}
&\colhead{$\lambda$}&\colhead{$\rm M^{DU}_{c,min}$}} 
\startdata
1& 0.88& 0.56& 0.037& -2.1& 1.6& 0.62& 0.10& 0.55 \nl 
2& 0.85& 0.58& 0.060& -2.1& 0.8& 0.72& 0.02& 0.56 \nl
\enddata
\tablecomments{Masses are in M$_{\sun}$}

\end{deluxetable}

\clearpage

\begin{deluxetable}{cc}
\tablewidth{33pc}
\tablenum{3}
\tablecaption{Adopted Prompt PN ejection parameters}
\tablehead{\colhead{Parameter}&\colhead{Value}}
\startdata
Luminosity& $4000\ L_{\sun}$\nl
Radius&  $400\ R_{\sun}$\nl
Mass& $0.58\ M_{\sun}$\nl
Core Mass& $0.54\ M_{\sun}$\nl
Mass-Loss Rate& $3.2\times 10^{-5}\,\msolyr$\nl
Time in Stage& 2000 yr\nl
\enddata
\tablecomments{The values in this table are estimated from Figure~2 of
BS88a for a 1.0 M$_{\sun}$, Z=0.001 model. All values are appropriate
between point C and the vertical dashed line on this figure. The
radius and luminosity are the estimated lower limits. The mass and
core mass are taken from their listed values. The mass-loss rate is
calculated from our mass-loss prescription. The time in this stage is
estimated from the BS88a graph.}
\end{deluxetable}

\clearpage

\plotone{k648modx.eps}

\clearpage

\plotone{k648abun.eps}

\end{document}